\documentclass{article}

\pdfoutput=1

\usepackage{lipsum}
\usepackage{amsmath}
\usepackage{amsfonts}
\usepackage{graphicx}
\usepackage{epstopdf}
\usepackage{amssymb}
\usepackage{scrextend}
\usepackage{booktabs}
\usepackage{algorithmic}
\usepackage{makecell}
\usepackage{placeins}
\usepackage{hyperref}
\usepackage[norule]{footmisc}
\usepackage{multirow}
\usepackage{cleveref}
\usepackage{dsfont}
\usepackage{amsopn}
\usepackage{diagbox}
\usepackage{mathtools}

\numberwithin{equation}{section}

\title{Representing Model Discrepancy in Bound-to-Bound Data Collaboration\thanks{
This work was funded by U.S. Department of Energy, National Nuclear Security Administration, under Award Number DE-NA0002375.}}

\author{
  Wenyu Li\footnotemark[2]
  \and
  Arun Hegde\footnotemark[2]
  \and
  James Oreluk\footnotemark[2]
  \and
  Andrew Packard\footnotemark[2]
  \and
  Michael Frenklach\thanks{Department of Mechanical Engineering, University of California at Berkeley, CA 94720-1740
    (wenyuli@berkeley.edu, arun.hegde@berkeley.edu, jim.oreluk@berkeley.edu, apackard@berkeley.edu, frenklach@berkeley.edu).}
}
\date{}

\begin{document}
\maketitle

% REQUIRED
\begin{abstract}
We extended the existing methodology in Bound-to-Bound Data Collaboration (B2BDC), an optimization-based deterministic uncertainty quantification (UQ) framework, to explicitly take into account model discrepancy. The discrepancy was represented as a linear combination of finite basis functions and the feasible set was constructed according to a collection of modified model-data constraints. Formulas for making predictions were also modified to include the model discrepancy function. Prior information about the model discrepancy can be added to the framework as additional constraints. Dataset consistency, a central feature of B2BDC, was generalized based on the extended framework.
\end{abstract}

\section{Introduction}
\label{sec:introduction}
During the past few decades, computational capabilities and data availability have seen substantial growth in many scientific and engineering fields. The growing demand for predictive models with quantifiable uncertainty has developed into an active research area, uncertainty quantification (UQ) \cite{oberkampf2010verification}. Two principal objectives of UQ are inference of model parameters, also known as the inverse or calibration problem utilizing a set of known data (the {\it training} set), and model prediction outside such a set. Theories and methods have been developed from both statistical and deterministic perspectives. In the statistical perspective, specifically under a Bayesian inference framework \cite{gelman2013bayesian}, the prior distribution of model parameters is updated by the likelihood of observations through Bayes' theorem. The produced posterior distribution is utilized for model parameter inference and predictions. In the deterministic perspective, for example the Bound-to-Bound Data Collaboration (B2BDC) method \cite{feeley2004consistency,frenklach2002prediction,frenklach2004collaborative,seiler2006numerical} employed in this paper, posterior ranges on model parameters and predictions are obtained through inequality-constrained optimization.

The two perspectives, while complementing each others in some aspects, address essentially the same problems and produce comparable outcomes. Both methodologies emphasize the critical role of identifying the source of uncertainty \cite{frenklach2016comparison}. In addition to Bayesian calibration methods, we also note that B2BDC shares conceptual similarities with Bayesian history matching \cite{craig1997pressure} and methods under the set membership framework \cite{chisci1996recursive,garulli2000conditional,milanese1991optimal,schweppe1968recursive}. Bayesian history matching retains a probabilistic interpretation of the data and defines a non-implausible region in the parameter space that contains all acceptable parameter vectors. Although both methods seek a region containing valid parameter vectors, Bayesian history matching uses stochastic emulators and improves the quality of the emulators through iterative updates of the non-implausible region, whereas B2BDC employs polynomial response surfaces and focuses primarily on evaluation of uncertainty in predictions \cite{frenklach2002prediction,frenklach2004collaborative}. Similar to the setup in B2BDC, the set-membership framework in robust control describes uncertainty in prior information and data by constraints. Methods developed under this framework also formulate estimation/prediction of quantities of interest (QOIs) as regions defined by inequality constraints or solutions to optimization problems \cite{milanese1991optimal,walter1990estimation}. However, these methods differ from B2BDC in that they often pursue a simply shaped approximation of the complex regions formed therein, e.g., a bounding ellipsoid \cite{fogel1982,schweppe1968recursive} or a minimum-volume bounding parallelotope \cite{chisci1996recursive}. B2BDC, on the other hand, uses polynomial surrogate models and handles the resulting nonconvexity through convex relaxation, leading to global guarantees on optimality \cite{seiler2006numerical}. A prior B2BDC analysis showed that approximation of the feasible set by a bounding ellipsoid or polytope may lead to overly conservative prediction results \cite{russi2010uncertainty}.

When analysis suggests disagreement among models and data, there are three possible causes: the model is correct and the data are flawed, the data are correct and the model is flawed, or both are flawed. In the present study, we focus on the second case, where we have more confidence in the data than the model. The first case has been a subject of our past studies \cite{feeley2004consistency,hegde2018consistency}, and the third case will be left as a challenge for future work. In the work of Kennedy and O'Hagan \cite{kennedy2001bayesian}, experimental observations are assumed to be noisy measurement of the underlying true process which represents reality,
\begin{equation} \label{eq:observation}
y = \mathcal{R}(s) + \epsilon,
\end{equation}
where $\epsilon$ is the measurement noise, $s$ are the scenario parameters, and $\mathcal{R}(s)$ represents reality. The scenario parameters are controllable properties known from the experimental setup, like the initial temperature and pressure, and can vary from experiment to experiment. In any scientific endeavor, knowledge of the true process is an idealization; a model, considered as tentatively entertained \cite{box1965experimental}, may have a systematic error in prediction. Kennedy and O'Hagan \cite{kennedy2001bayesian} suggested to describe the uncertainty in the model form as an additive term, $\delta$, referred to as model {\it inadequacy}, to the model output,
\begin{equation} \label{eq:model discrepancy KO}
\mathcal{R}(s) = M(x^*,s) + \delta(s),
\end{equation}
\noindent where $x^*$ are the underlying true calibration parameters. The model parameters are uncertain parameters intrinsic to the model, $M(\cdot)$, and share common values across all experiments. 

This approach of compensating for model discrepancy has received substantial interest and following (see, e.g., \cite{bayarri2007framework,brynjarsdottir2014learning,higdon2004combining,joseph2009statistical,kleiber2013parameter,plumlee2017bayesian,qian2008bayesian,storlie2015calibration,wang2009bayesian}), some using a Gaussian process (GP) \cite{sacks1989design,stein2012interpolation} to represent $\delta(s)$ \cite{brynjarsdottir2014learning,higdon2004combining,plumlee2017bayesian,qian2008bayesian,storlie2015calibration,wang2009bayesian} and others a functional decomposition \cite{joseph2009statistical,kleiber2013parameter}, while referring to model inadequacy as model discrepancy, model bias, model form uncertainty, model error, and model form error. Efforts have also been made to overcome the difficulty in identifying model discrepancy and model parameters individually, and to improve prediction performance at conditions different from the training data. For example, Brynjarsd{\'o}ttir and O'Hagan \cite{brynjarsdottir2014learning} put constraints on the GP realization of model discrepancy at specific conditions derived from domain knowledge. Plumlee \cite{plumlee2017bayesian} argued that the prior distribution of model discrepancy should be orthogonal to the gradient of the model under certain assumptions. Wang et al. \cite{wang2009bayesian} estimated the model discrepancy and model parameters separately. Joseph and Melkote \cite{joseph2009statistical} constructed a statistical model of discrepancy in a sequential manner, limiting its contribution to the prediction. In Bayesian history matching, model discrepancy is taken into account by including a structured term in its statistical model of the relation between data and simulator's output \cite{goldstein2009reified}.

Our objective here is to resolve disagreement among models and data in the deterministic setting of B2BDC using the perspective of \Cref{eq:model discrepancy KO}. The optimization-based framework of B2BDC represents uncertainty by sets and has been successfully applied in several domains, including combustion science \cite{feeley2004consistency,frenklach2007,frenklach2004collaborative,iavarone2018,russi2008,dlr2017} and engineering \cite{utah}, atmospheric chemistry \cite{smith2006}, quantum chemistry \cite{edwards2014}, and system biology \cite{feeley2006,feeleyThesis,yi2005}. In the present work, we expand the B2BDC formalism by adding a deterministic model-form discrepancy function to the constraints derived using model and data, conceptually following Kennedy and O'Hagan \cite{kennedy2001bayesian}. We start with a brief recount of B2BDC in \Cref{sec:w/o model inadequacy}, followed by reformulating the feasible set and prediction problems with model discrepancy in \Cref{sec:w/ model inadequacy}. Application of the proposed methodology is presented in \Cref{sec:result} for two examples, a simple mass-spring-damper and a more realistic combustion system. Further interpretation of model discrepancy as a general consistency measure is discussed in \Cref{sec:consistency}. We conclude with summarizing comments in \Cref{sec:conclusion}.

\section{Methodology} 
\label{sec:methodology}
Bound-to-Bound Data Collaboration (B2BDC) is a deterministic optimization-based framework for systematically combining models and experimental data with quantified uncertainties \cite{feeley2004consistency,frenklach2002prediction,frenklach2004collaborative,seiler2006numerical}. In this framework, the uncertainties in experimental data are specified by intervals $[L_e, U_e]$, where $L_e$ and $U_e$ are the lower and upper bounds assessed for the $e$-th quantity of interest (QOI). A prior uncertainty region of model parameters derived from domain knowledge is also given and denoted by $\mathcal{H}$. The collection of all provided information is referred to as a {\it dataset}. 

\subsection{B2BDC without model discrepancy}
\label{sec:w/o model inadequacy}
The experimental data are utilized to carve out a smaller region in $\mathcal{H}$, referred to as the {\it feasible set} and denoted by $\mathcal{F}$,
\begin{equation} 
\label{eq:feasible set}
\mathcal{F} = \{x \, | \, x\in \mathcal{H}, \, L_e \leq M(x,s_e) \leq U_e, \ e=1,2,\ldots,N \}.
\end{equation}
The constraints $L_e \leq M(x,s_e) \leq U_e$ are referred to as {\it model-data constraints} since they are derived by connecting model outputs with experimental bounds. The feasible set constitutes the posterior region of the model parameters which satisfy all model-data constraints. The dataset is referred to as {\it consistent} if its feasible set is non-empty and {\it inconsistent} otherwise. The feasibility is determined by calculating a numerical measure termed the {\it scalar consistency measure} (SCM) \cite{feeley2004consistency}. The quantity is defined by
\begin{equation}\label{eq:SCM}
\begin{aligned}
C_{\textrm{D}} \coloneqq \,& \underset{x\in\mathcal{H}}{\text{maximize}}
& & \gamma \\
& \text{subject to}
& & L_e + \gamma (\frac{U_e-L_e}{2}) \leq M_e(x) \leq U_e-\gamma(\frac{U_e-L_e}{2}), \\
& & & e=1,2,\ldots,N.
\end{aligned}
\end{equation}
The dataset is consistent if $C_{\rm{D}}\geq 0$ as its non-negative value proves the existence of a parameter vector that satisfies all the constraints. For a consistent dataset, the predicted interval $[L_p, \, U_p]$ for an unmeasured QOI can be calculated by solving
\begin{equation}
\label{eq:pred.min.w/o}
L_p  = \min_{x\in\mathcal{F}} M(x,s_p),
\end{equation}
\noindent and
\begin{equation}
\label{eq:pred.max.w/o}
U_p  = \max_{x\in\mathcal{F}} M(x,s_p).
\end{equation}
where $s_p$ are the corresponding scenario parameters.

An inconsistent dataset implies that the models, experimental data, and prior information are fundamentally incompatible with each other, making prediction essentially meaningless. In prior work, inconsistency was resolved by identifying likely offending experimental data. This was accomplished by computing sensitivities of the consistency measure \cite{feeley2004consistency} and/or minimal relaxations of the bounds to recover consistency \cite{hegde2018consistency}. In the present work, we focus on regaining dataset consistency and the ability to make predictions through scenario-dependent model corrections.

In practice, the optimization problems in \Cref{eq:SCM,eq:pred.min.w/o,eq:pred.max.w/o} are often challenging to solve numerically because the underlying model is not convex and expensive to evaluate. The B2BDC approach is to bracket the solution to the optimization problems with two quantities, referred to as the inner bound and outer bound, whose computation is more tractable \cite{feeleyThesis,seiler2006numerical}. Computation of the outer bound, in particular, motivates the use of a quadratic/polynomial or rational quadratic surrogate model for the underlying model at each scenario condition, i.e., $M(x,s_e)$. The outer bound is then obtained by solving a semidefinite program \cite{seiler2006numerical}. The inner bound is a local solution to the original optimization problem found by nonlinear constrained optimization solvers. In our B2BDC toolbox \cite{B2BDCgithub}, the inner bound is calculated with MATLAB function \texttt{fmincon} and the outer bound using SDP solver SeDuMi \cite{sturm1999using} or CVX package \cite{gb08,cvx}. Considering that many QOIs depend mainly on a small fraction of the model parameters, a phenomenon termed effective sparsity \cite{box1985some}, a subset of the model parameters that have a significant effect on $M(x,s_e)$, referred to as active parameters, is selected by a screening sensitivity analysis \cite{frenklach2007optimization}. The coefficients of the surrogate model are computed by fitting simulations of $M(x,s_e)$ at selected points, referred to as design points, in the active parameter space. Development of the active parameters results in a reduced computational cost of building and evaluating the surrogate models. In a typical B2BDC application, generation of surrogate models usually consumes the most CPU time. In addition to the savings provided by selecting active parameters for each QOI, the cost can be further reduced by running simulations at design points in parallel. Previous applications \cite{dlr2017,smith2011gri} showed the computation of inner and outer bounds took at most a few minutes on a desktop computer for datasets consisted of 55 and 102 model parameters along with 167 and 77 QOIs, respectively.

\subsection{B2BDC with model discrepancy}
\label{sec:w/ model inadequacy}
In practice, one may encounter an inconsistent dataset where the experimental data are more reliable than the model form (e.g., as in \cite{oreluk2018diagnostics}). In such a case, resolution of the inconsistency may be sought by compensating the difference between the model and data through introduction of a discrepancy function. We adopt the definition of model discrepancy proposed by Kennedy and O'Hagan in \Cref{eq:model discrepancy KO} and on that basis introduce a scenario-dependent \textit{discrepancy function} $\delta(s)$ into the model-data constraints,
\begin{equation} 
\label{eq:B2B with model discrepancy}
L_e \leq M(x,s_e) + \delta(s_e) \leq U_e, \ e=1,2,\ldots,N.
\end{equation}
We assume that the discrepancy takes the form of a linear combination of $n$ basis functions, $\{\Phi_i\}_{i=1}^n$,
\begin{equation} 
\label{eq:model discrepancy decomposition}
\delta(s) = \sum_{i=1}^n c_i \Phi_i(s),
\end{equation}
where $\{c_i\}_{i=1}^n$ are unknown coefficients, and $n=0$ refers hereafter to a zero discrepancy function (i.e., $\delta=0$). Similar forms of the discrepancy function have also been used by others \cite{joseph2009statistical}. Substitution of \Cref{eq:model discrepancy decomposition} into \Cref{eq:B2B with model discrepancy} results in,
\begin{equation} 
\label{eq:model-data constraint with model discrepancy}
L_e \leq M(x,s_e) + \sum_{i=1}^n c_i \Phi_i(s_e) \leq U_e, \ e=1,2,\ldots,N.
\end{equation}
The linear form in \Cref{eq:model discrepancy decomposition} is motivated by the fact that the existing tools in B2BDC are directly applicable with \Cref{eq:model-data constraint with model discrepancy} in the extended parameter space $(x, \, c)$: while $\{\Phi_i\}_{i=1}^n$ can be any set of nonlinear functions, the modified model-data constraints are linear in $\{c_i\}_{i=1}^n$. Representing model discrepancy by a linear combination of basis functions has been used by others \cite{joseph2009statistical}. However, our method is different from those in that it does not set as an objective to fit a particular model discrepancy function. Hence, orthogonality among basis functions, usually imposed to improve regression performance, is not required.

We define the joint feasible set in the extended parameter space of $x$ and $c$ by combining the prior uncertainty and modified model-data constraints,
\begin{equation} 
\label{eq:joint feasible set}
\mathcal{F}_\delta = \{(x, c)\, | \, x\in \mathcal{H}, \, c \in \mathcal{H}_{c}, \,  L_e \leq M(x,s_e) + \sum_{i=1}^n c_i \Phi_i(s_e) \leq U_e, \ e=1,2,\ldots,N\},
\end{equation}
where $\mathcal{H}_c$ represents the prior uncertainty region of the discrepancy-function coefficients. The feasibility of $\mathcal{F}_\delta$ can be calculated by modifying the model-data constraints accordingly in the SCM formula in \Cref{eq:SCM}. The projection of $\mathcal{F}_\delta$ on the model parameter space is,
\begin{equation} 
\label{eq:modified feasible set}
\widetilde{\mathcal{F}} = \{x \, | \,  \exists \hat{c}: (x,\hat{c}) \in \mathcal{F}_\delta \},
\end{equation}
which represents the set of feasible model parameters after including the discrepancy function. When the joint feasible set is not empty, prediction at an unmeasured scenario $s_p$ can be obtained by solving the modified versions of \Cref{eq:pred.min.w/o,eq:pred.max.w/o}:
\begin{equation}
\label{eq:modified pred.min.w/o}
L_p  = \min_{(x,c) \in \mathcal{F}_\delta} M(x,s_p)+\sum_{i=1}^n c_i \Phi_i(s_p),
\end{equation}
\noindent and
\begin{equation}
\label{eq:modified pred.max.w/o}
U_p  = \max_{(x,c) \in \mathcal{F}_\delta} M(x,s_p)+\sum_{i=1}^n c_i \Phi_i(s_p).
\end{equation}

The proposed framework treats the data as the sum of model output and discrepancy function as implied by \Cref{eq:model discrepancy KO}. The joint feasible set, defined over the extended parameter space $(x, \, c)$, therefore represents only combinations of the model and discrepancy function that are consistent with the data. It is not possible to elucidate the model or the discrepancy function separately without further assumptions and/or additional information.

A challenge with the above approach is the choice of basis functions. A practitioner may simply choose a set with the least number of basis functions that resolves dataset inconsistency, following for instance the Akaike information criterion \cite{akaike1974new} of penalizing larger number of parameters, and thus prevent overfitting. If extra insight is provided, various forms of discrepancy function can be investigated before making the final decision based on considerations in addition to the requirement that dataset consistency is recovered.

The developed framework with model discrepancy expressed using \Cref{eq:model discrepancy decomposition} has a general feature that, for a given dataset and a prediction QOI, the prediction interval becomes systematically wider if additional basis functions are included. To understand this, suppose two sets of basis functions are used in an analysis, with the second being a superset of the first. Let vector $c$ represent the coefficients for the shared basis functions $\{\Phi_i\}_{i=1}^n$ and $c'$ the coefficient vector for the additional basis functions $\{\Phi'_j\}_{j=1}^{n'}$, i.e.,
\begin{equation}
\begin{aligned}    
& \delta^1 (s) = \sum_{i=1}^n c_i \Phi_i (s) \\
& \delta^2 (s) = \delta^1(s) + \sum_{j=1}^{n'} c_j' \Phi'_j (s).
\end{aligned}    
\end{equation}
The corresponding joint feasible sets formed by \Cref{eq:joint feasible set} are denoted by  $\mathcal{F}_\delta^1$ and $\mathcal{F}_\delta^2$. Any feasible point $(x,\,c) \in \mathcal{F}_\delta^1$ is also feasible for $\mathcal{F}_\delta^2$ by setting $c'$ to zero, i.e., $(x,\,c,\,c')|_{c'=0} \in \mathcal{F}_\delta^2$. Therefore, the posterior uncertainty interval of the QOI, predicted on $\mathcal{F}_\delta^1$, is always contained by that predicted on $\mathcal{F}_\delta^2$. The increased uncertainty in the prediction interval can depend on the prediction QOI, the dataset, and the selected basis functions, as will be demonstrated in \Cref{sec:result}.

%\subsection{Prior and posterior of model discrepancy} \label{sec:prior and posterior on model discrepancy}
Previous work (e.g., \cite{brynjarsdottir2014learning,plumlee2017bayesian}) has demonstrated the value of including prior knowledge of the model discrepancy function when applying statistical UQ methods. In B2BDC, this can be accomplished by incorporating additional constraints. For example, sign constraints on the discrepancy function, or its derivatives, can be enforced at specified scenario conditions by introducing linear inequalities in $c$. An example of forcing model discrepancy function to be positive at selected scenarios is
\begin{equation}\label{eq:positive_constraint_on_model_discrepancy}
    \sum_{i=1}^n c_i\phi_i(s_j) > 0, \ j=1,2,\ldots.
\end{equation}
The effect of such constraints is automatically propagated to predictions through augmenting the feasibility constraint in \Cref{eq:modified pred.min.w/o,eq:modified pred.max.w/o}. Another example of constraining the magnitude of model discrepancy function is given in \Cref{sec:spring additional constraint}.

The posterior uncertainty of the model discrepancy function at any specified scenario $s_p$ can be calculated by solving the prediction problems in \Cref{eq:modified pred.min.w/o,eq:modified pred.max.w/o} with the objective replaced by $\delta(s_p)$, i.e.,
\begin{equation}
\label{eq:pred.min.delta}
L_\delta  = \min_{(x,c) \in \mathcal{F}_\delta} \sum_{i=1}^n c_i \Phi_i(s_p),
\end{equation}
\noindent and
\begin{equation}
\label{eq:pred.max.delta}
U_\delta  = \max_{(x,c) \in \mathcal{F}_\delta} \sum_{i=1}^n c_i \Phi_i(s_p).
\end{equation}
Repeating this computation at various conditions in the scenario parameter space can identify regions where uncertainty in the model discrepancy function is large. In a similar manner, the posterior uncertainty can be calculated for each discrepancy-function coefficient $c_i$.

\section{Numerical examples} \label{sec:result}
Application of the extended B2BDC framework is demonstrated with two examples, an illustrative mass-spring-damper system and a realistic hydrogen combustion system. In each example, we started with a postulated ``true'' model to represent the underlying true process. An inadequate model for the analysis was then created by omitting some parts of the true model and a true calibration parameter value was selected. The developed framework was applied to the inadequate model and the following results are reported and discussed for different choices of basis functions.

\begin{enumerate}
\item Dataset consistency 
\item If the dataset is consistent, a) the predicted intervals at interpolated and extrapolated conditions; b) whether the true process values are contained in the predicted intervals, and to a secondary point, whether the true calibration parameter $x^*$ is in the feasible set.
\end{enumerate}
\vspace{3mm}

The computation was conducted for two levels of experimental uncertainty to provide a more comprehensive characterization of the developed method. To clarify the nomenclature, $\delta^*$ and $\delta$ are used to represent the true model discrepancy defined in \Cref{eq:model discrepancy KO} and the linear combination in \Cref{eq:model discrepancy decomposition}, respectively. We also differentiate between \textit{interpolated} and \textit{extrapolated} predictions, where the former refers to $s_p$ lying within the training domain and the latter outside.

All the example scripts along with the general B2BDC software \cite{B2BDCgithub}---i.e., everything required to reproduce the results reported in this paper---can be found using the GitHub link \url{https://github.com/B2BDC/}.

\subsection{One-dimensional mass-spring-damper system}
\label{sec:spring system}
\begin{figure}[!htb]
\centering
\includegraphics[width=\linewidth]{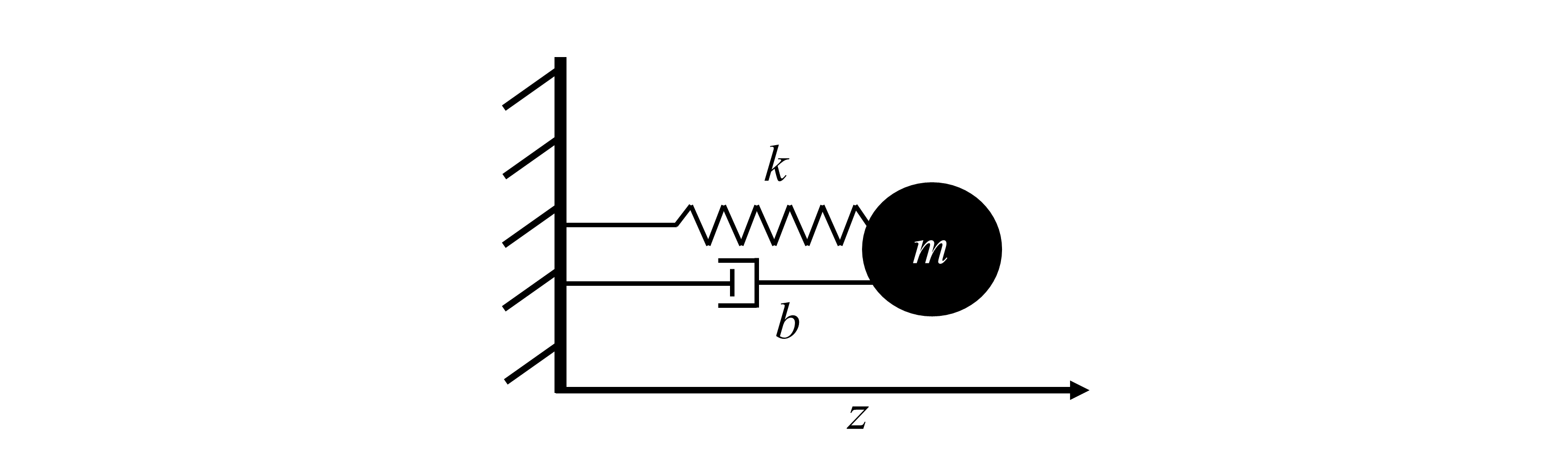}
\caption{Sketch plot of the mass-spring-damper system.}
\label{fig:spring system}
\end{figure}

\begin{figure}[!htb]
\centering
\includegraphics[width=\linewidth]{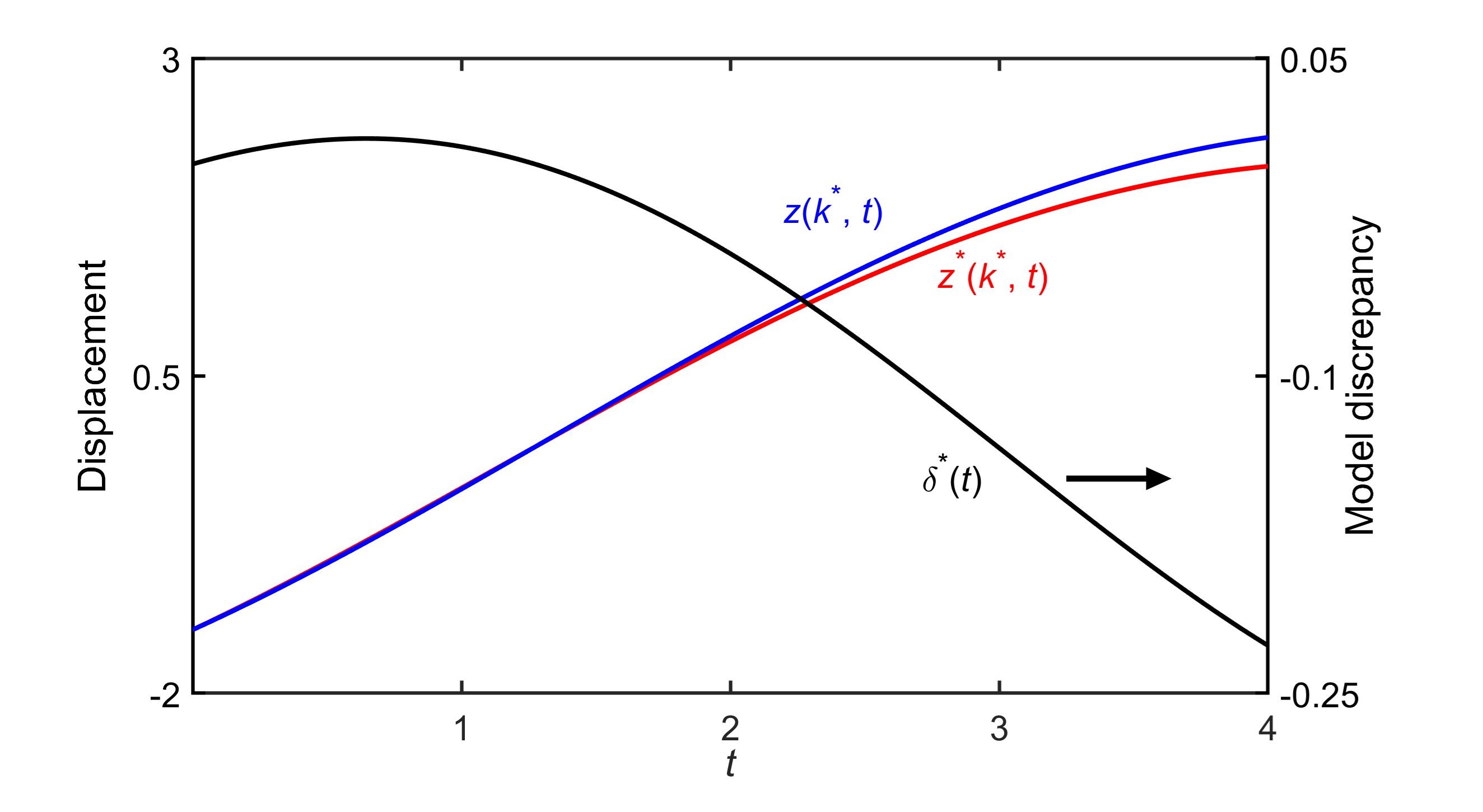}
\caption{The true model solution $z^*(k^*,t)$, the inadequate model solution $z(k^*,t)$ and the model discrepancy function $\delta^*(t) = z^*(k^*,t)-z(k^*,t)$.}
\label{fig:z* z and delta}
\end{figure}

The force, $F$, needed to extend or compress a spring by a small distance, $z$, is expressed using Hooke's law
\begin{equation} \label{eq:Hooke's law}
    F = -k z,
\end{equation}
where $k$ is a constant characteristic of the spring, its stiffness. We now consider a simple system: a ball attached to a spring, whose other end is fixed at a wall, sketched in \Cref{fig:spring system}.
The ball has a mass $m = 1$ and is placed initially at $z_0 = -1.5$ with an initial velocity $v_0 = 1$. In addition to the force exerted by the spring, the motion of the ball is also affected by a damping force proportional to the ball's velocity. Thus, the evolution of the ball's displacement is described by
\begin{equation} \label{eq:spring evolution}
\begin{aligned}
&\frac{d^2 z}{dt} = -kz - b \frac{dz}{dt}, \\
&\left.z\right\vert_{t=0} = z_0 = -1.5, \\
&\left.\frac{dz}{dt}\right\vert_{t=0} = v_0 = 1,
\end{aligned}
\end{equation}
where $b$ is the constant coefficient of the damping force and its value was set to 0.05. For a given $k$, displacement evolution of system described by \Cref{eq:spring evolution}---the ``true'' model in this example---is the solution to a second order, constant coefficient, ordinary differential equation and has analytic form
\begin{equation} \label{eq:true spring displacement}
    z^*(k,t) = e^{-bt/2} \left[ \frac{v_0+0.5bz_0}{\sqrt{k-b^2/4}}\sin \left(\sqrt{k-b^2/4}\ t\right)
    + z_0 \cos\left(\sqrt{k-b^2/4}\ t\right) \right].
\end{equation}
The ``inadequate'' model was constructed by neglecting the damping force (i.e., $b = 0$), which results in the solution
\begin{equation} \label{eq:given spring displacement}
    z(k,t) = \frac{v_0}{\sqrt{k}}\sin\left(\sqrt{k}\ t\right)+z_0 \cos(\sqrt{k}\ t).
\end{equation}
In both the true and inadequate models, the stiffness $k$ is the model parameter and the time $t$ is the scenario parameter. The true stiffness of the spring --- the true calibration parameter value --- was selected to be $k^*=0.25$ with the prior uncertainty interval $\mathcal{H} = [0.2,\, 0.3]$. The real displacement was evaluated with $z^*(k^*,t)$. The displacements computed by the two models with $k = k^*$ and their difference, the model discrepancy defined in \Cref{eq:model discrepancy KO}, are demonstrated in \Cref{fig:z* z and delta} for $t\in[0, \,4]$.

The QOIs for this example were chosen to be the displacements of the ball at specified times $t_e$. The dataset was composed of twenty of these QOIs in the scenario region $t\in[0, \, 3]$. For each QOI, an ``observed'' value was generated by adding uniform noise with a prescribed maximum magnitude $\epsilon$ to the true process value,
\begin{equation} 
\label{eq:spring data}
\begin{aligned}
    &z_e = z^*(k^*,t_e) + \epsilon\,u_e, \\
    &u_e \sim \mathcal{U}(-1,\,1), \quad e=1,2,\ldots,20.
\end{aligned}
\end{equation}
QOI uncertainty bounds were generated by setting $L_e = z_e-\epsilon$ and $U_e = z_e+\epsilon$. The present analysis was performed with $\epsilon$ values of 0.05 and 0.1. Three prediction QOIs were generated for $t$ values of 1.5, 3.2, and 4. The first prediction case occurs at a scenario within the training-set domain of $[0,\, 3]$ and is an interpolated prediction. The second and third cases occur at scenarios outside the training-set domain and are extrapolated predictions.
\FloatBarrier

\subsubsection{Dataset consistency and QOI prediction} \label{sec:spring consistency and prediction}
\begin{figure}[!htb]
\centering
\includegraphics[width=\linewidth]{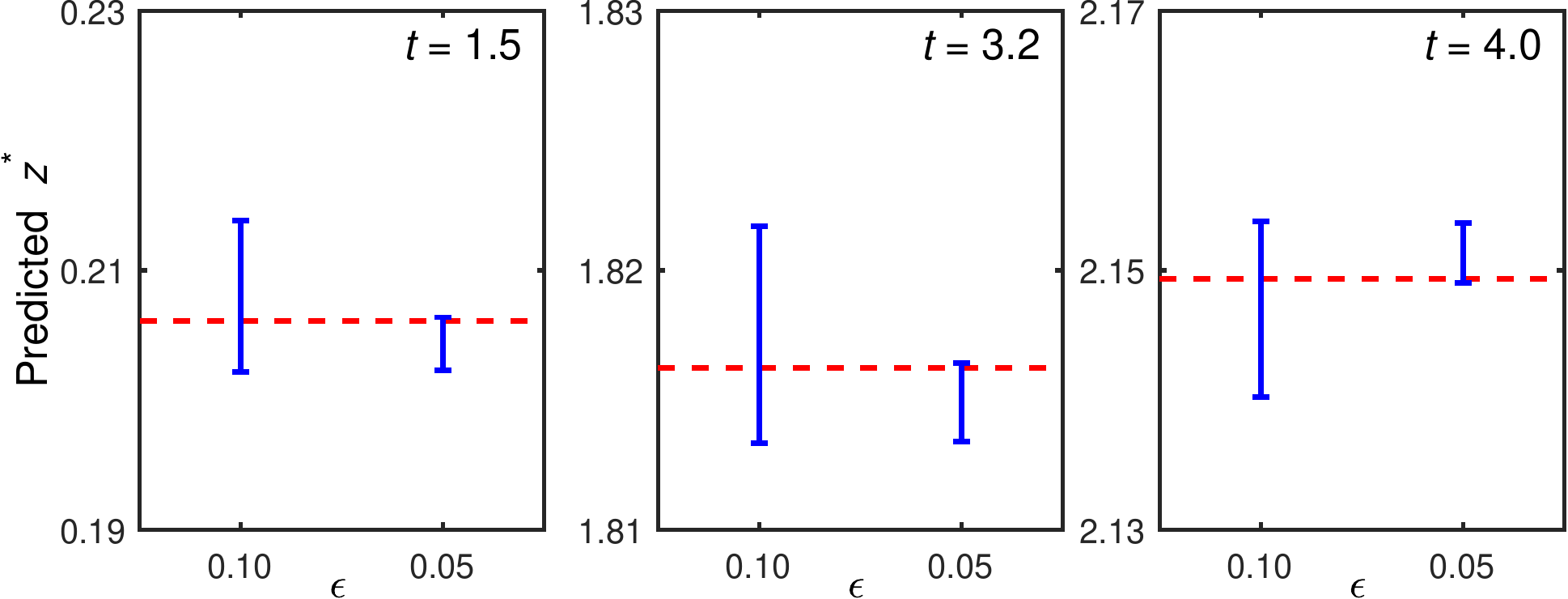}
\caption{Computed QOI prediction intervals for the mass-spring-damper example using the true model. The horizontal red dashed lines mark the displacement computed with the true model and true calibration parameter value, $z^*(k^*,t)$. The vertical blue solid lines designate the B2BDC predicted intervals, computed by solving optimization problems in \Cref{eq:pred.min.w/o,eq:pred.max.w/o}.}
\label{fig:spring prediction true}
\end{figure}

We first considered the ideal situation where the true model, given by \Cref{eq:true spring displacement}, and the formulas in \Cref{sec:w/o model inadequacy} were used in the B2BDC calculations. The prediction results are displayed in \Cref{fig:spring prediction true}. With this setup, the dataset is consistent with $k^*$ being feasible and the predicted intervals contain the true process values for both tested $\epsilon$'s. The length of the predicted intervals at each prediction scenario is shorter for a smaller value of $\epsilon$, indicating that more accurate measurements produce more accurate predictions.

We then moved to a more realistic situation where an inadequate model, given by \Cref{eq:given spring displacement}, was examined through the modified B2BDC framework, described in \Cref{sec:w/ model inadequacy}. Four different model discrepancy functions,
\begin{equation} 
\label{eq:polynomial}
\delta(t) = \sum\limits_{i=1}^{n} c_{i-1} t^{i-1}, \quad  n =1,2,3,4,
\end{equation}
were tested in addition to the case where $\delta=0$. The discrepancy function is a polynomial in $t$ of degree $n-1$.

The outcome of the dataset consistency analysis is summarized in \Cref{tab:spring consistency}. Examination of these results shows that the dataset is inconsistent for both values of $\epsilon$ when $n = 0$, i.e., a zero model discrepancy function was used. For $\epsilon = 0.05$, a quadratic $\delta$ is required to obtain dataset consistency. In this case, $k^*$ is also found to be feasible. For $\epsilon = 0.1$, a constant $\delta$ is enough to achieve consistency. However, $k^*$ becomes feasible only after using a linear $\delta$.

\begin{table}[!htb]
\caption{Results of the dataset consistency analysis}\label{tab:spring consistency}
\begin{center}
\begin{tabular}{c c c c c c } \hline
\multirow{2}{*}{ $\epsilon$} & \multicolumn{5}{c}{$n$} \\ \cline{2-6}
 & 0 & 1 & 2 & 3 & 4 \\ \hline
 0.05 & inconsistent & inconsistent & inconsistent & $k^*\in\widetilde{\mathcal{F}}$ & $k^*\in\widetilde{\mathcal{F}}$ \\ 
 0.10 & inconsistent & $k^*\not\in\widetilde{\mathcal{F}}$ & $k^*\in\widetilde{\mathcal{F}}$ & $k^*\in\widetilde{\mathcal{F}}$ & $k^*\in\widetilde{\mathcal{F}}$ \\ \hline
\end{tabular}
\end{center}
\end{table}

\begin{figure}[!htb]
%% \centering
\includegraphics[width=\linewidth]{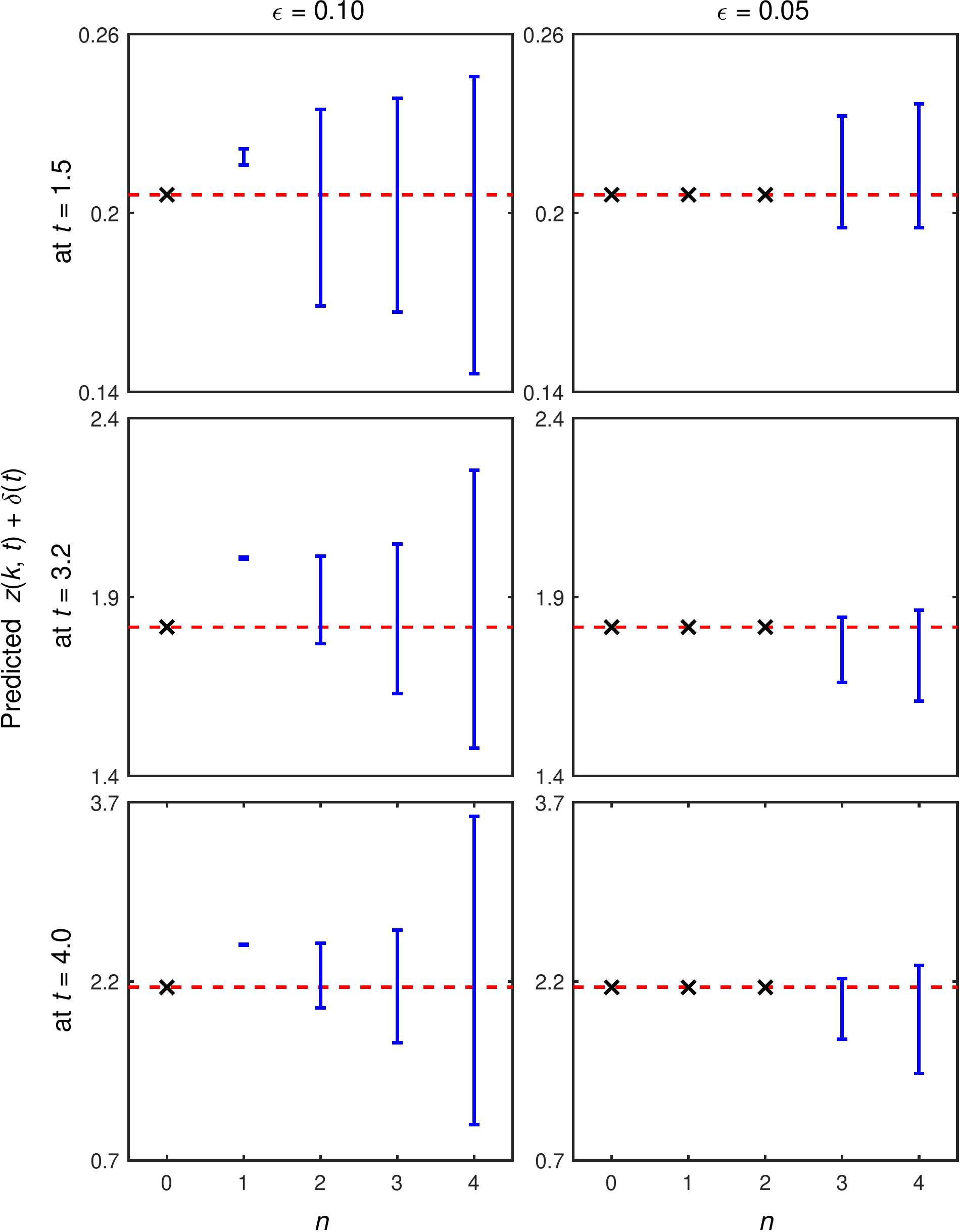}
\caption{Predicted QOI intervals at t = 1.5, 3.2 and 4. The horizontal red dashed lines are the displacement derived with the true model and evaluated at the true model parameter value, $z^*(k^*,t)$. The vertical blue solid lines designate the B2BDC predicted intervals, computed by solving the optimization problems in \Cref{eq:modified pred.min.w/o,eq:modified pred.max.w/o}. The $\times$'s mark dataset inconsistency.}
\label{fig:pred QOI}
\end{figure}

The predicted QOI intervals are displayed in \Cref{fig:pred QOI} for $t=1.5$, $3.2$, and $4$. As expected, the prediction intervals with a higher order $\delta$ are wider for both $\epsilon$ values. In the cases where $\delta$ produced a consistent dataset for both $\epsilon$ values, a shorter prediction interval is observed with the smaller $\epsilon$. For $\epsilon=0.1$, the QOI interval predicted using a constant $\delta$ does not contain the true value at all time instances. With a linear $\delta$, the predicted interval contains the true value at all three time instances. The predicted interval contains the true value for all tested times and for both values of $\epsilon$ with a quadratic and cubic $\delta$. 
\FloatBarrier

\subsubsection{Posterior bounds of model parameter and discrepancy-function coefficients} \label{sec:spring posterior bounds}
We now examine the posterior uncertainty bounds of model parameter $k$ and discrepancy-function coefficients $\{c_i\}_{i=1}^n$ obtained for different polynomial orders of $\delta$. These bounds are the 1-dimensional projection of the joint feasible set $\mathcal{F}_\delta$ onto coordinate directions. The volume ratio of the joint feasible set and the multidimensional box, whose sides are the posterior projections of the parameters, was calculated as the fraction of $10^6$ samples uniformly distributed in the box, that lay in $\mathcal{F}_\delta$. The results are presented in \Cref{tab:posterior uncertainty range}. For comparison, the posterior interval of $k$ obtained with the true model is also listed. The computed volume ratio results show that the joint feasible set becomes progressively smaller relative to the box as dimension increases.
\begin{table}[!htb]
\caption{Projection of the joint feasible set, computed with the inadequate model defined in \Cref{eq:given spring displacement} and different model discrepancy functions defined in \Cref{eq:polynomial}, onto coordinate directions of model parameter $k$ and discrepancy-function coefficients $\{c_i\}_{i=1}^n$, as well as computed volume ratio of the joint feasible set to the box made of the projected intervals. The symbol $\varnothing$ represents an empty posterior uncertainty due to dataset inconsistency.}\label{tab:posterior uncertainty range}
\begin{footnotesize}
\begin{center}
\begin{tabular}{l c c c c c c} \hline
\multicolumn{1}{c}{$n$} & $k$ & $c_0$ & $c_1$ & $c_2$ & $c_3$ & Volume ratio \\\cmidrule(r){1-1} \cmidrule(rl){2-6} \cmidrule(l){7-7}
\multicolumn{7}{c}{$\epsilon = 0.05$} \\
 & [0.24, 0.25]$^*$ & & & & \\
0 & $\varnothing$ & & & & & \\
1 & $\varnothing$ & $\varnothing$ & & & &\\
2 & $\varnothing$ & $\varnothing$ & $\varnothing$ & & &\\
3 & [0.20, 0.30] & [-0.02, 0.03] & [-0.08, 0.06] &  [-0.05, 0.00] & & $5.2\times 10^{-3}$\\
4 & [0.20, 0.30] & [-0.02, 0.03] & [-0.12, 0.06] &  [-0.13, 0.14] & [-0.05, 0.03] & $2.3\times 10^{-4}$\\
\rule{0pt}{2mm} \\
\multicolumn{7}{c}{$\epsilon = 0.10$} \\
 & [0.24, 0.26]$^*$ & & & & \\
0 &  $\varnothing$ & & & & & \\
1 & [0.20, 0.21] & [0.00, 0.01] &  &  & & $5.2\times 10^{-1}$\\ 
2 & [0.20, 0.30] & [-0.01, 0.04] & [-0.11, 0.01] &  &  & $4.9\times 10^{-2}$\\ 
3 & [0.20, 0.30] & [-0.02, 0.04] & [-0.13, 0.06] &  [-0.05, 0.03] & & $1.2\times 10^{-2}$ \\ 
4 & [0.20, 0.30] & [-0.04, 0.05] & [-0.19, 0.19] &  [-0.31, 0.22] & [-0.07, 0.09] & $2.0\times 10^{-4}$\\ \hline
\end{tabular}
\end{center}
\end{footnotesize}
\begin{addmargin}[12mm]{10mm}
\footnotesize{$^*$Posterior uncertainty interval obtained with the true model.} \\
\end{addmargin}
\end{table}

The B2BDC analysis with the true model resulted in a significantly narrower posterior uncertainty interval for model parameter $k$ as compared to its prior; the interval in this case, by design, contains the true calibration parameter value. With the inadequate model and a constant model discrepancy function ($n=1$) at $\epsilon=0.1$, an even narrower posterior interval was obtained; however, the true value, $k^*$, is completely missed. With a higher order $\delta$, the posterior interval covers the same range as the prior. This outcome can be explained by considering two factors that affected the posterior interval of $k$.

\begin{figure}[!htb]
\centering
\includegraphics[width=\linewidth]{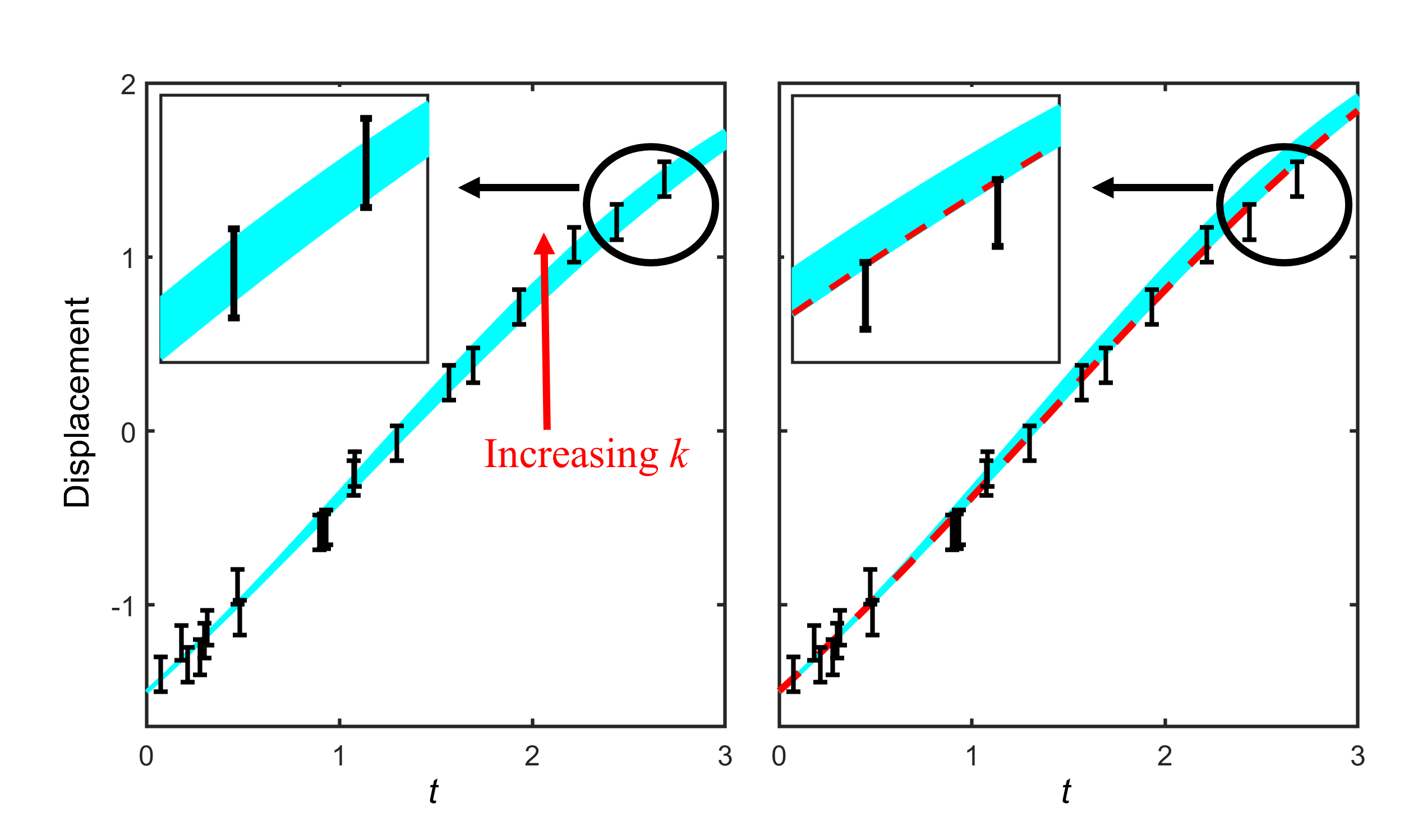}
\caption{Displacements computed with the true (left) and the inadequate (right) models for various $k$ values drawn from its prior interval $[0.2,\,0.3]$ (cyan regions). The black vertical bars are observation QOI bounds. The red dashed line is one feasible realization of $z(k,t)+c_0$ with $k=0.2$ and $c_0=0.005$. The insets are zoomed on the last two observations for $t\in[2.3,\, 2.8]$.}
\label{fig:QOI span}
\end{figure}

Firstly, the inadequate model has a different functional dependency on the model parameter $k$, resulting in a problem specific change of the posterior bounds: feasible $k$ values for the true model can become infeasible for the inadequate model and vice versa. In the current example, this can be visually shown by comparing the displacement predicted using the true and inadequate models and its dependency on model parameter $k$, as demonstrated in \Cref{fig:QOI span}. Plotted in this figure are $z^*(k,t)$ and $z(k,t)$ computed for different $k$'s drawn from its prior interval along with the experimental bounds for the case where $\epsilon=0.1$. For given $t$, larger $k$ values produced larger displacements for both models. The resulting displacement bands (shown in cyan) cover similar vertical regions at smaller $t$ values but the band for the inadequate model gradually shifts upward with increasing magnitude at larger $t$ values comparing to that for the true model. For the last two observations, shown in the right inset plot, only a small portion of the band satisfies the QOI bounds. Note that this portion corresponds to smaller $k$ values. However, predictions with these smaller $k$ values invalidated at least one other QOI bound, motivating the use of $\delta$ to resolve inconsistency.

The second factor, as discussed in \Cref{sec:w/ model inadequacy}, is that inclusion of a higher order $\delta$ always results in a wider posterior interval. For a constant $\delta$ at $\epsilon = 0.1$, the posterior interval widened from the empty set (a zero $\delta$) to an interval with finite length. With the constant $\delta$, feasible $(k,c_0)$ can be found with $k$ limited to a very small region close to the prior lower bound. The red dashed curve in \Cref{fig:QOI span} corresponds to the prediction with one of the feasible $(k,c_0)$.

\begin{figure}[!htb]
\centering
\includegraphics[width=\linewidth]{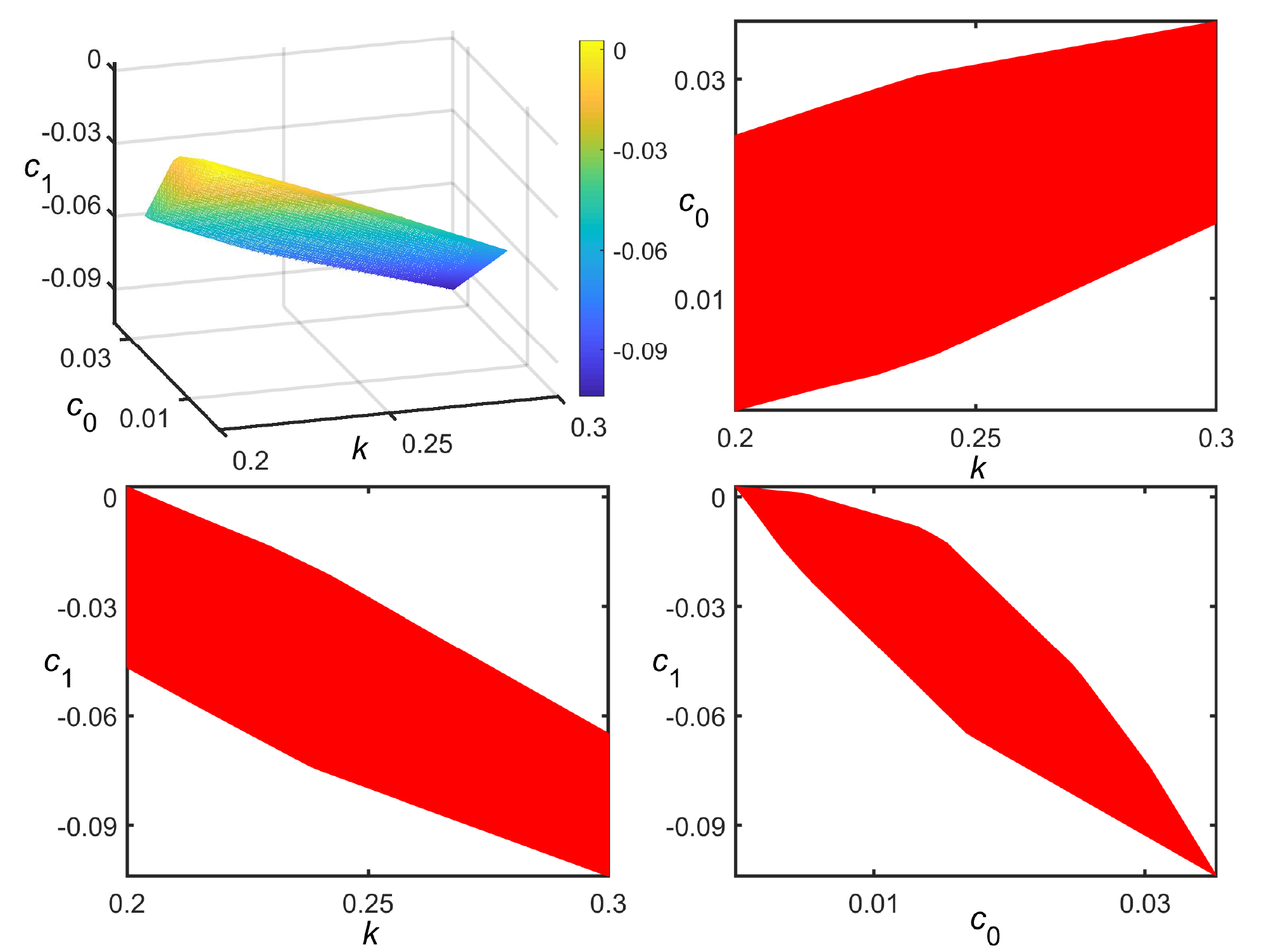}
\caption{Joint feasible set of $k$, $c_0$, and $c_1$ and its 2-dimensional projections (colored in red), computed with linear $\delta$ and $\epsilon=0.1$. The color bar in the upper-left figure color codes the values of $c_1$. The axes' limits of $k$, $c_0$, and $c_1$ were set to their calculated posterior uncertainty bounds.}
\label{fig:confounding}
\end{figure}

The posterior uncertainty intervals of $\{c_i\}_{i=1}^n$ also become systematically wider with a higher polynomial order of $\delta$, as expected. The enlarged posterior uncertainty intervals associated with individual parameters $k$ and $\{c_i\}_{i=1}^n$ are related to the phenomenon usually referred to in statistical literature (e.g., \cite{brynjarsdottir2014learning}) as confounding, manifesting itself in the presence of a strong correlation between model parameter(s) and model discrepancy despite their relatively wider marginal posterior distributions. We demonstrate this from a deterministic perspective by the plots shown in \Cref{fig:confounding}, generated for the case of a linear $\delta$ at $\epsilon=0.1$. The plots display the joint feasible set of $k$, $c_0$ and $c_1$ along with its 2-dimensional projections. The three-dimensional plot clearly shows that the joint feasible set occupies only a small fraction of the enclosing cube. Inspection of the projections indicates that at a fixed $k$ value, the uncertainty in $c_0$ and $c_1$ is reduced, on average, to 66 and 46\%, of their posterior ranges, and at fixed $c_0$, the uncertainty in $c_1$ is reduced, on average, to 26\% of its posterior range.
\FloatBarrier
\subsubsection{Posterior uncertainty of model discrepancy} \label{sec:spring model discrepancy posterior}
\begin{figure}[!htb]
\centering
\includegraphics[width=\linewidth]{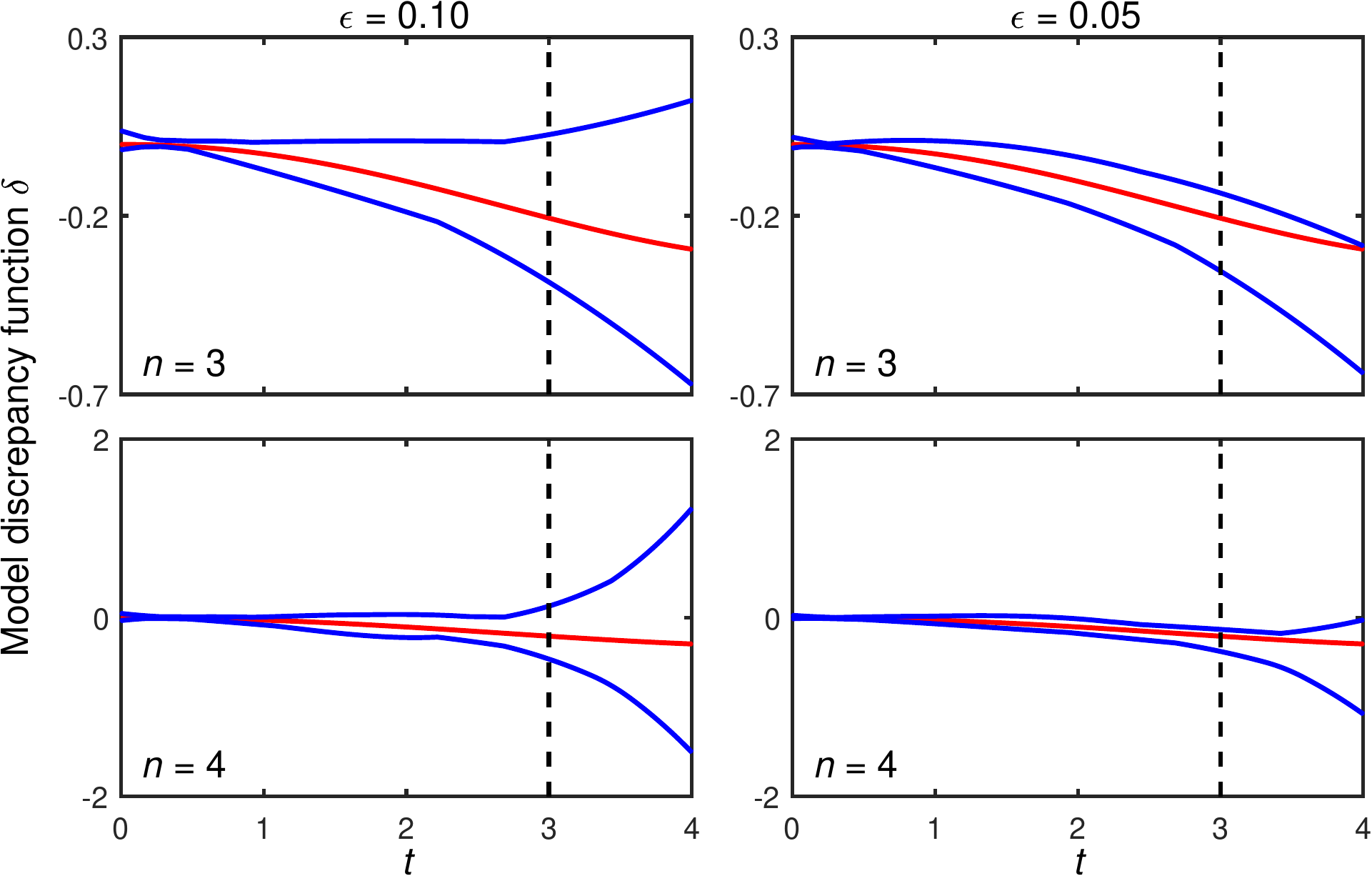}
\caption{Uncertainty bounds of quadratic and cubic $\delta$ (blue lines). The red line is the true model discrepancy $\delta^*$.}
\label{fig:delta interval}
\end{figure}

We now examine the upper and lower bounds of $\delta$ predicted at 1000 discrete time points, $t_i$, equally spaced in $[0,\,4]$. The bounds were calculated by solving problems in \Cref{eq:pred.min.delta,eq:pred.max.delta}. This region is divided into the interpolation zone ($t\in[0,\, 3]$), where data exists, and the extrapolation zone ($t\in[3,\, 4]$) for comparison. The uncertainty bands are shown in \Cref{fig:delta interval} for quadratic and cubic $\delta$; they were generated by linearly interpolating adjacent upper and lower bounds. 

Inspection of these results shows that the computed uncertainty bounds enclose $\delta^*$ in both the interpolation and extrapolation zones for both quadratic and cubic $\delta$. The width of the predicted uncertainty band is effectively constrained within the interpolation zone. The predicted uncertainty band starts to widen toward the end of the interpolation zone and diverges rapidly in the extrapolation zone. The observed divergence is more dramatic for a cubic $\delta$ than a quadratic $\delta$. The uncertainty band for a fixed $\delta$ is overall narrower with a smaller $\epsilon$ in both interpolation and extrapolation zones.
\FloatBarrier
\subsubsection{Additional constraints on model discrepancy}
\label{sec:spring additional constraint}
As discussed in \Cref{sec:w/ model inadequacy}, constraints derived from domain knowledge about the model discrepancy can be included in the B2BDC calculations. We illustrate this feature with the following example.

\begin{figure}[!htb]
\centering
\includegraphics[width=\linewidth]{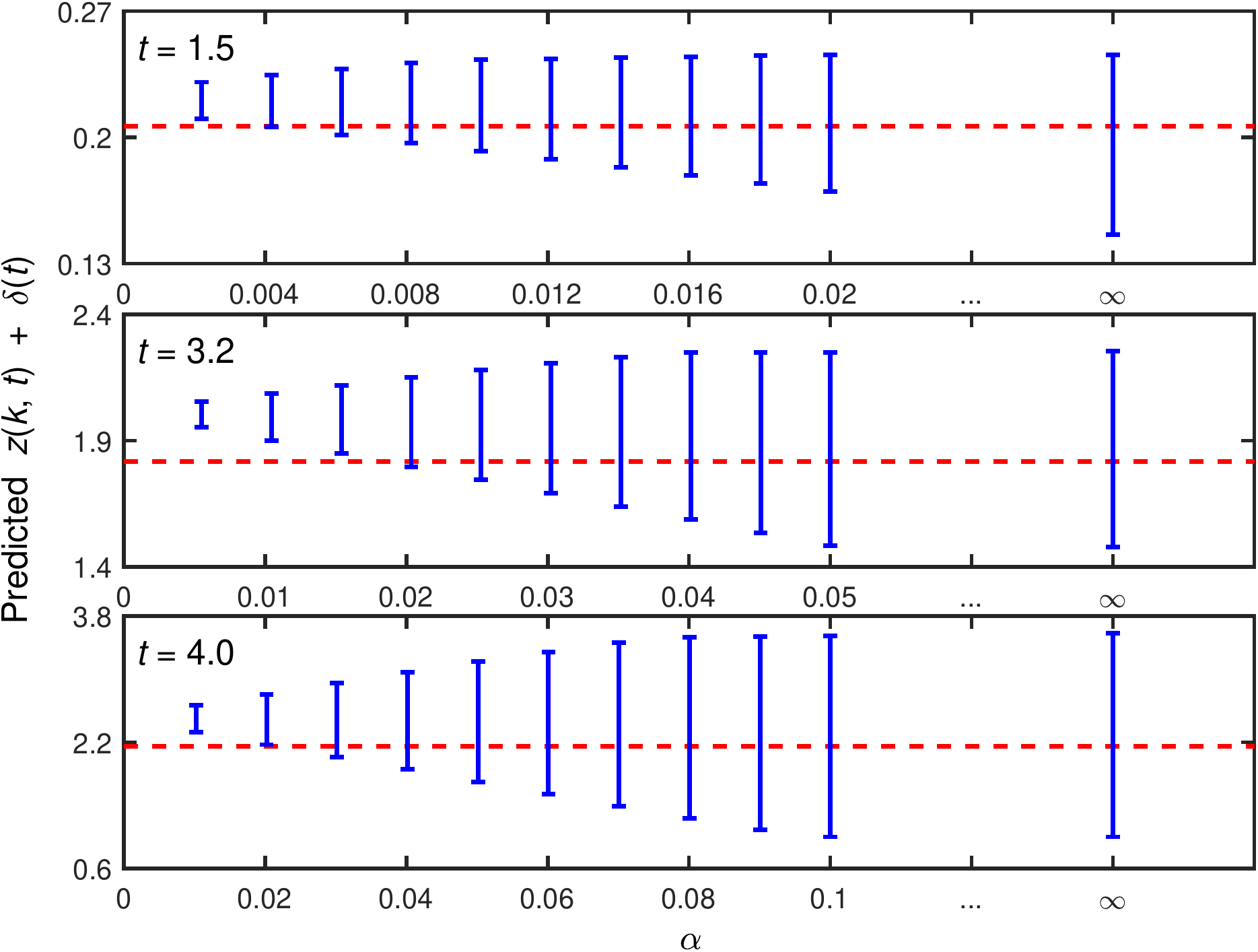}
\caption{Interpolation and extrapolation intervals computed by solving problems in \Cref{eq:modified pred.min.w/o,eq:modified pred.max.w/o} with the extra constraint in \Cref{eq:constrain MD} for cubic $\delta$ at $\epsilon=0.1$. The red dashed lines are the true prediction values.}
\label{fig:constrain B}
\end{figure}

Let us assume that although we introduced a discrepancy function, we would still like to rely on the model more than on the introduced correction when making predictions. This idea reflects the general spirit of some existing work in the literature (e.g., \cite{joseph2009statistical}). This requirement can be attained by selecting among all feasible values of $\delta$ those that have their magnitude, averaged over data and prediction scenarios, below a prescribed threshold, $\alpha$,
\begin{equation} \label{eq:constrain MD}
\frac{1}{N+1} (\left\lvert \delta(t_p) \right\rvert + \sum_{i=1}^N \left\lvert \delta(t_i) \right\rvert )  \leq \alpha,
\end{equation}
where $N$ is the number of experimental QOIs. The constraint was added to the joint feasible set constructed using \Cref{eq:joint feasible set} and predictions were made with varying values of $\alpha$. The results for $\epsilon=0.1$ and cubic $\delta$ are shown in \Cref{fig:constrain B}.
As expected, the predicted interval increases for larger $\alpha$, reaching the value obtained without using \cref{eq:constrain MD} eventually, as this additional constraint becomes inactive.
\FloatBarrier

\subsection{A hydrogen combustion model}
\label{sec:H2 combustion}
In this section we apply the formalism described above to a hydrogen combustion model: a homogeneous adiabatic H\textsubscript{2}-air reaction system at constant volume. The evolution of the system states, i.e., species concentrations and temperature, is simulated numerically by solving a set of ordinary differential equations. The time derivatives of species concentrations and temperature are calculated based on the specified chemical reaction mechanism and the energy equation. Simulations with detailed (21 reactions \cite{you2012process}) and reduced (5 reactions \cite{williams2008detailed}) mechanisms, listed in \Cref{append:mechanisms}, were considered as the true and inadequate models, respectively. The model parameters, denoted by $\lambda \in \mathbb{R}^5$, are logarithm of the multipliers associated with the five rate constants shared by both mechanisms, with their prior uncertainties taken from \cite{you2012process}. The true calibration parameter value was specified as $\lambda^* = \mathbf{0}$, where $\mathbf{0}$ is a vector of zeros. 

The normalized scenario parameters, $s_1$, $s_2$ and $s_3$, were defined as
\begin{equation} \label{eq:normalized scenario}
\begin{aligned}
    s_1 = &\frac{1000/T-1000/T_{\text{center}}}{1000/T_{\text{low}} - 1000/T_{\text{high}}}, \\ &  T_{\text{center}}=1370 \ \text{K}, \ T_{\text{low}} = 1200 \ \text{K}, \ T_{\text{high}} = 1600 \ \text{K}, \\
     s_2 = &\frac{\ln{P}-\ln{P_{\text{center}}}}{\ln{P_{\text{high}}} - \ln{P_{\text{low}}}}, \\ & P_{\text{center}}=3.2\ \text{atm}, \ P_{\text{low}} = 1\ \text{atm}, \ T_{\text{high}} = 10\ \text{atm}, \\
     s_3 = &\frac{\phi-\phi_{\text{center}}}{\phi_{\text{high}} - \phi_{\text{low}}}, \\ 
    & \phi_{\text{center}}=1, \ \phi_{\text{low}} = 0.75, \ \phi_{\text{high}} = 1.25,
\end{aligned}
\end{equation}
where $T$, $P$ and $\phi$ are the initial temperature, initial pressure and equivalence ratio of the mixture, respectively. In this example, equivalence ratio is the ratio of hydrogen to oxygen concentrations in the initial mixture to that in the stoichiometric mixture. The use of inverse temperature and logarithm of pressure for defining $s_1$ and $s_2$ are common in the combustion field, e.g., \cite{williams2008detailed}. 

A dataset was constructed using a second-order orthogonal design \cite{myers2009response} over the scenario region $[-1,\,1]^3$. The scenario parameter values are listed in \Cref{tab:orthogonal design}.
\begin{table}[!htb]
\caption{Design conditions for the training data. }\label{tab:orthogonal design}
\begin{center}
\begin{tabular}{c c c c c c c} \hline
Design index & $s_1$ & $T$ (K) & $s_2$ & $P$ (atm) & $s_3$ & $\phi$ \\ \hline
1 & 1 & 1200 & 1 & 10 & 1 & 1.25 \\ 
2 & 1& 1200 & 1 & 10 & -1 & 0.75 \\ 
3 & 1& 1200 & -1&  1 & 1& 1.25 \\ 
4 & 1 & 1200 & -1& 1 & -1& 0.75 \\
5 & -1& 1600 & 1& 10 & 1& 1.25 \\ 
6 & -1& 1600 & 1& 10 & -1& 0.75 \\ 
7 & -1& 1600 & -1& 1 & 1& 1.25 \\ 
8 & -1& 1600 & -1& 1 & -1& 0.75 \\ 
9 & 0 & 1370 & 0 & 3.2 & 0 & 1 \\ 
10 & 1.215 & 1170 & 0 & 3.2 & 0 & 1 \\ 
11 & -1.215 & 1660 & 0 & 3.2 & 0 & 1 \\ 
12 & 0 & 1370 & 1.215 & 12.8 & 0 & 1 \\ 
13 & 0 & 1370 & -1.215 & 0.78 & 0 & 1 \\ 
14 & 0 & 1370 & 0 & 3.2 & 1.215 & 1.3 \\
15 & 0 & 1370 & 0 & 3.2 & -1.215 & 0.7 \\ \hline
\end{tabular}
\end{center}
\end{table}
For each of the scenario conditions, the corresponding QOI was defined as the time when the hydrogen concentration drops to half of its initial value. This QOI was computed numerically from the simulated hydrogen concentration profile and denoted by $t_{1/2}^*(\lambda,T,P,\phi)$ and $t_{1/2}(\lambda,T,P,\phi)$ for the true and inadequate models, respectively. ``Measured'' QOIs, denoted by $t_i$, were generated by adding a relative noise to the true process values, specified as $t_{1/2}^*$ evaluated at the true calibration parameter value $\lambda^*$,
\begin{equation} \label{eq:t observation}
\begin{aligned}
    &t_i = t_{1/2}^*(\lambda^*,T_i,P_i,\phi_i)(1 + \epsilon\,u_i), \\
    &u_i \sim \mathcal{U}(-1,\,1), \quad i=1,2,\ldots,15.
\end{aligned}
\end{equation}
The maximum noise magnitude, $\epsilon$, was assigned values of 0.01 and 0.005. As before, the uncertainty bounds were generated by computing $[(1-\epsilon)t_i, (1+\epsilon)t_i]$. The QOI computed with the inadequate model has no analytic solution and a quadratic surrogate model $S_i$ was generated for each QOI such that $S_i(\lambda) \approx {\rm{ln}}(t_{1/2}(\lambda, T_i,P_i,\phi_i))$. As in the previous example, we consider a polynomial model discrepancy function (\Cref{tab:delta in H2}), but now with the scenario parameters $s_1$, $s_2$ and $s_3$ defined by \Cref{eq:normalized scenario}.
\begin{table}[!htb]
\caption{Tested model discrepancy functions.}\label{tab:delta in H2}
\begin{center}
\begin{tabular}{l c} \hline
\makecell[c]{Model discrepancy} & Number of basis function $n$ \\ \hline
No $\delta$ & 0 \rule[-1.2ex]{0pt}{2.6 ex}\\
$\delta = c_0$ & 1 \rule[-1.2ex]{0pt}{0pt}\\
$\delta = c_0 + \sum_{i=1}^3 c_is_i$ & 4 \rule[-1.2ex]{0pt}{0pt} \\
$\delta = c_0 + \sum_{i=1}^3 c_is_i + \sum_{i,j=1;\, i \leq j}^3 c_{i,j}s_is_j$  & 10 \rule[-1.3ex]{0pt}{0pt} \\
\hline
\end{tabular}
\end{center}
\end{table}

\subsubsection{Dataset consistency and QOI prediction}
\label{sec:hydrogen dataset consistency and prediction}
Dataset consistency was calculated first and the results are given in \Cref{tab:inference result}.
\begin{table}[!htb]
\caption{Results of dataset consistency and the distance between true model parameter value $\lambda^*$ and the feasible set.}\label{tab:inference result}
\begin{center}
\begin{tabular}{c c c} \hline
$n$ & Dataset consistency & $d_{\lambda^*}$ \\ \hline
& \multicolumn{2}{c}{$\epsilon=0.01$} \\
0 & Inconsistent & ---  \\ 
1 &Inconsistent & --- \\ 
4 & Consistent& 0.167  \\ 
10 & Consistent & 0.047  \\
\rule{0pt}{2mm} \\
& \multicolumn{2}{c}{$\epsilon=0.005$} \\ 
0 & Inconsistent & ---  \\ 
1 &Inconsistent & --- \\ 
4 & Inconsistent& ---  \\ 
10 & Consistent & 0.072  \\ \hline
\end{tabular}
\end{center}
\end{table}
Inspection of these results shows that with $\epsilon=0.01$, the dataset is inconsistent for both the zero $\delta$ and constant $\delta$ cases, and becomes consistent when linear and quadratic $\delta$ are used. After $\epsilon$ was lowered to 0.005, linear $\delta$ is insufficient to keep the dataset consistent. For cases where the dataset is consistent, the distance between the true calibration parameter value and the feasible set, denoted by $d_{\lambda^*}$ and defined as
\begin{equation}
\label{eq:distance from true model parameter}
d_{\lambda^*}^2  = \min_{(\lambda,c) \in \mathcal{F}_\delta} (\lambda-\lambda^*)^T(\lambda-\lambda^*),
\end{equation}
was calculated and the results are also reported in \Cref{tab:inference result}. In all these cases, the true calibration parameter value is not in the feasible set. Its distance from the feasible set is larger when lower order $\delta$ or smaller $\epsilon$ were used.

\begin{figure}[!htb]
\centering
\includegraphics[width=\linewidth]{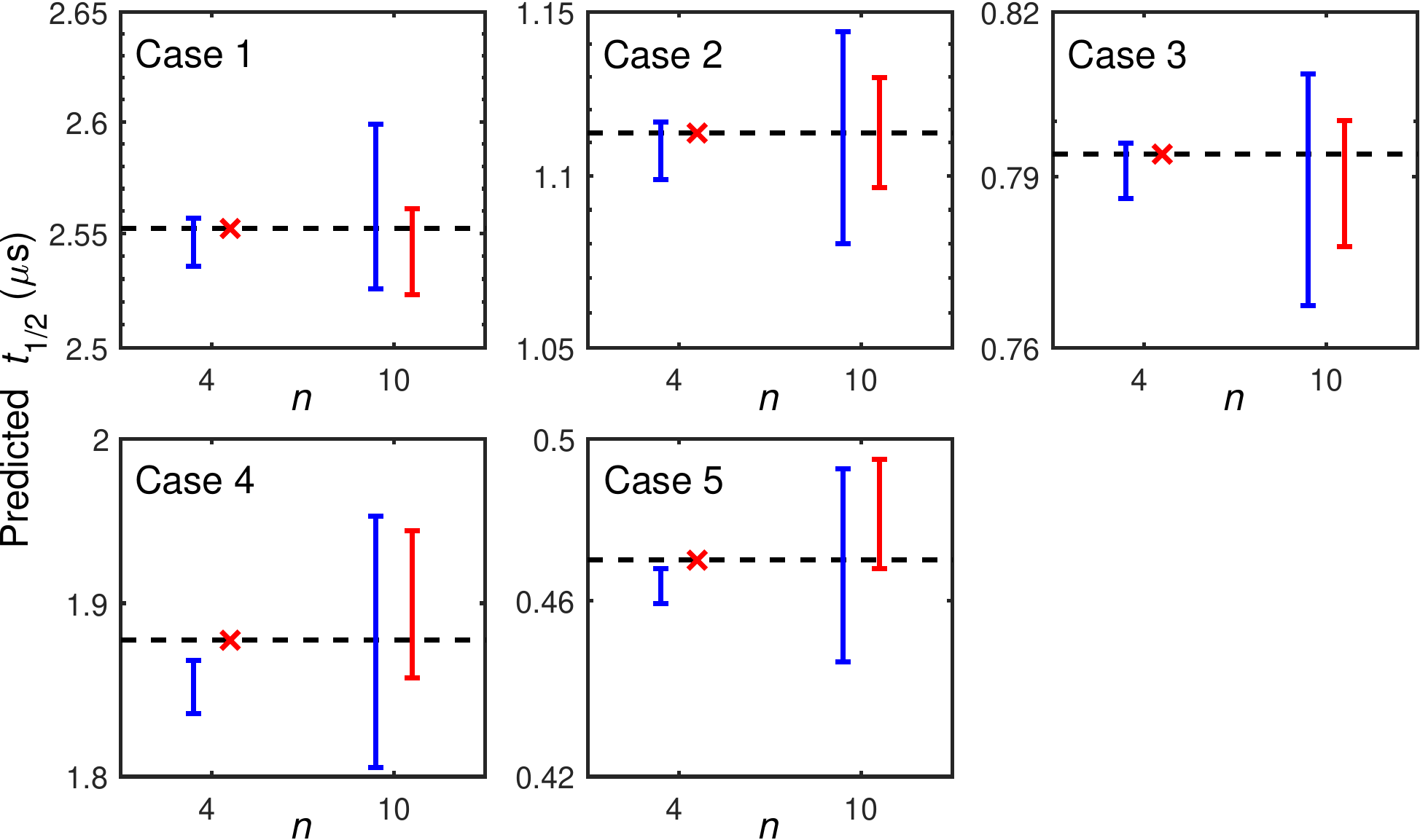}
\caption{QOI prediction intervals for the five cases in \Cref{tab:interpolate and extrapolate}. The black dashed lines are the true QOI values. The predicted QOI intervals are drawn as blue ($\epsilon=0.01$) and red ($\epsilon=0.005$) vertical lines. The red $\times$'s mark dataset inconsistency.}
\label{fig:H2 prediction}
\end{figure}

For cases where the dataset is consistent, model predictions were computed at one interpolated and four extrapolated scenarios, which are specified in \Cref{tab:interpolate and extrapolate}.
\begin{table}[!htb]
\caption{Scenario parameter values for model prediction.}\label{tab:interpolate and extrapolate}
\begin{center}
\begin{tabular}{c c c c c c c c} \hline
Case index & Prediction & $s_1$ &  $T$ (K) & $s_2$ & $P$ (atm) & $s_3$ & $\phi$ \\ \hline
1& Interpolation & -0.6 & 1500 & 0.4 & 5 & 0 & 1 \\ 
2& Extrapolation  & -1.67 &  1800 & 0.4 & 5 & 0 & 1 \\ 
3& Extrapolation  & -1 & 1600 & 1.16 &  12 & 0 & 1 \\ 
4& Extrapolation  & -1 & 1600 & 0.4 & 5 & 1.6 & 1.4 \\ 
5& Extrapolation  & -1.67 & 1800 & 1.16 & 12 & 1.6 & 1.4 \\ \hline
\end{tabular}
\end{center}
\end{table}
The results are depicted in \Cref{fig:H2 prediction}.
Again, the lengths of predicted intervals are shorter with linear $\delta$ as compared to quadratic $\delta$. Similarly, smaller values of $\epsilon$ produced shorter predictions. At $\epsilon=0.01$, the predicted interval with a linear $\delta$ contains the true value for cases 1, 2 and 3, but underpredicts the target for cases 4 and 5. With a quadratic $\delta$, the predicted intervals contain the true values for all tested cases at both $\epsilon$ values.

\subsubsection{Inference of model discrepancy}
\label{sec:hydrogen infer model discrepancy}
The projection of the feasible set on the parameter space of discrepancy function coefficients, $c$'s, describes not one but a set of discrepancy functions that are consistent with the data. The following analysis with the linear $\delta$ and $\epsilon=0.01$ shows an example of inferring model discrepancy from B2BDC calculations. The posterior uncertainty bounds of the discrepancy-function coefficients were calculated and the results are given in \Cref{tab:inference about linear MI}. The volume ratio of the joint feasible set to the multidimensional box, specified similarly as in \Cref{sec:spring system}, is $2.6\times10^{-8}$ based on $10^9$ samples.
\begin{table}[!htb]
\caption{Projection of the joint feasible set, computed with linear $\delta$ at $\epsilon=0.01$, onto coordinate directions of each discrepancy-function coefficient.}\label{tab:inference about linear MI}
\begin{center}
\begin{tabular}{c c } \hline
Coefficient & Posterior uncertainty bounds  \\ \cmidrule(lr){1-1} \cmidrule(lr){2-2}
$c_0$ & [-0.139, 0.112]  \\
$c_1$ & [-0.018, 0.005] \\ 
$c_2$ & [-0.015, -0.006]  \\ 
$c_3$ & [-0.042, -0.013]  \\ \hline
\end{tabular}
\end{center}
\end{table}
The results show that all feasible $c_2$ and $c_3$ are negative since the calculated posterior upper bounds are negative for these two coefficients. For the linear $\delta$, the coefficients are also the partial derivatives of the discrepancy function with respect to the scenario parameters, i.e.,
\begin{equation} \label{eq:partial_derivative_of_delta}
    c_i = \frac{\partial \delta}{\partial s_i}.
\end{equation}
All feasible $\delta$'s are therefore smaller at larger $s_2$ or $s_3$ values given other scenario parameters fixed.

\begin{figure}[!htb]
\centering
\includegraphics[width=\linewidth]{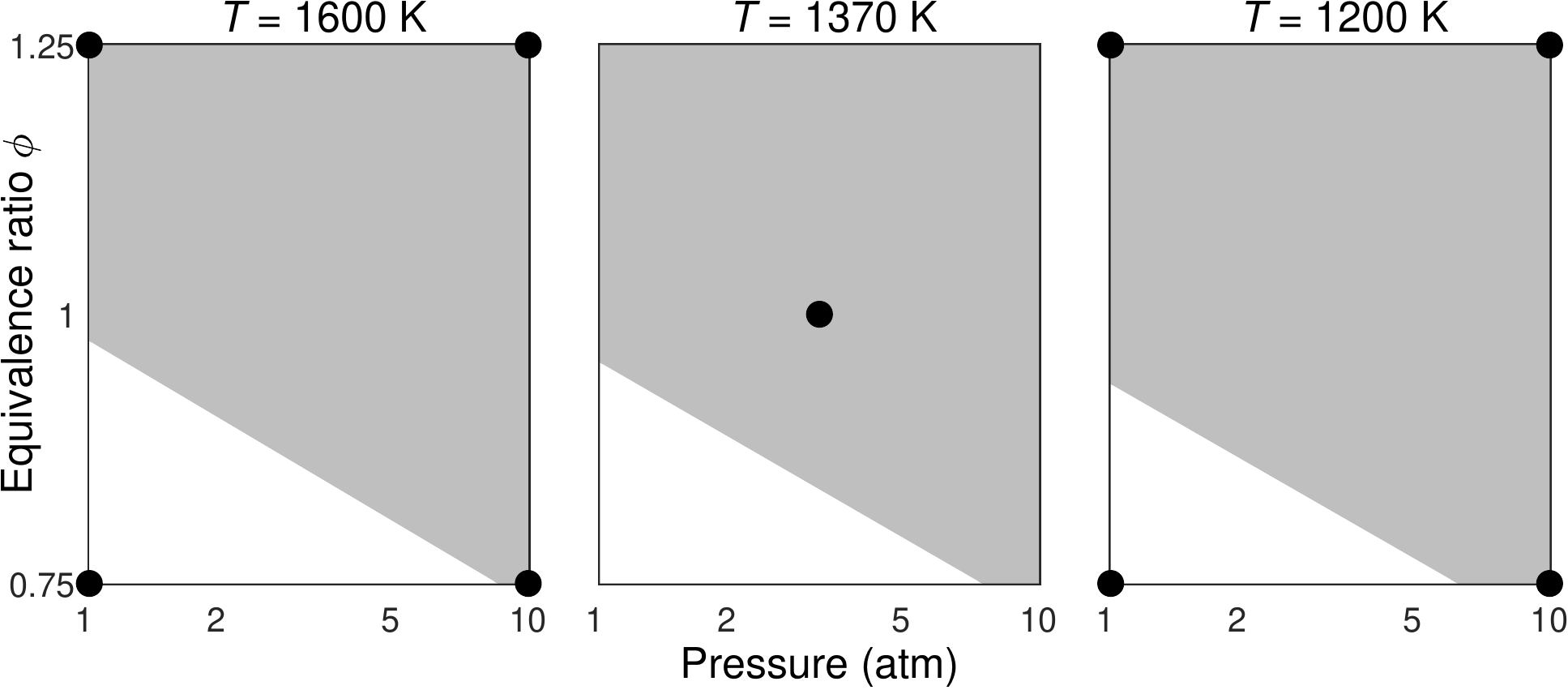}
\caption{The sign of model discrepancy function in the scenario region $(s_2,\, s_3) \in [-1,\,1]^2$ for temperature values 1600, 1370 and 1200 K. The grey region represents scenarios where $U_\delta<0$ and the white one indicates where the interval $[L_\delta,\, U_\delta]$ contains 0. The black points are design scenarios in \Cref{tab:orthogonal design}.}
\label{fig:delta sign}
\end{figure}

The predicted interval of $\delta$, i.e., $[L_\delta,\, U_\delta]$ from \Cref{eq:pred.min.delta,eq:pred.max.delta}, was then calculated in the $s_2$-$s_3$ ($P$-$\phi$) space at three fixed $s_1$ ($T$) values. The computed intervals were examined to determine the sign of feasible $\delta$'s at each specified scenario and the results are shown in \Cref{fig:delta sign}.
Similar patterns are observed for the three tested temperature values: except for a lower left triangle region where both pressure and equivalence ratio are relatively small, all feasible $\delta$'s are negative. As a result of dataset consistency, predictions made at the grey-region scenarios always add a negative correction to the model output, suggesting that the inadequate model systematically overpredicts the QOI. Combining the results that $c_2$ and $c_3$ are negative in the feasible set, the overprediction is likely to be stronger at larger $s_2$ and $s_3$ values, i.e., at higher pressures and equivalence ratios.

\section{Discrepancy as a consistency measure} \label{sec:consistency}
The above examples and discussion primarily focus on the impact of model discrepancy on prediction. The inclusion of discrepancy into B2BDC also provides the opportunity to calculate a more general consistency measure. For a given collection of basis functions $\{\Phi_i\}_{i=1}^{n}$, define this consistency measure as
\begin{equation}
\label{eq:generalcm}
\begin{aligned}
&\min_{x,c} \quad f(c) \\
& \ \ \text{s.t.} \quad L_e \leq M_e(x) + \sum_{i=1}^{n} c_i \Phi_i(s_e) \leq U_e \\
& \qquad \quad x \in \mathcal{H}\\
& \qquad \quad c \in \mathcal{H}_c \\
& \qquad \quad e=1,..,N,
\end{aligned}
\end{equation}
where the objective $f(\cdot)$ is a function of only the coefficients $\{c_i\}_{i=1}^n$ and reflects the ``complexity'' of the discrepancy function. In essence, \Cref{eq:generalcm} asks the following question: what is the least complex discrepancy function required to recover dataset consistency? Different choices of $f(\cdot)$ and $\Phi_i(\cdot)$ produce different consistency measures. For example, defining the complexity and discrepancy functions as
\begin{equation}
\begin{aligned}
& f(c)= \sum_{i=1}^N |c_i| \\
&\Phi_i(s) = -\mathds{1}_{\{s_i\}}(s) =
\begin{cases} -1 & s = s_i \\
0 & \text{otherwise} \end{cases} 
\end{aligned}
\end{equation}
where ${n} = N$ and the $\{s_i\}_{i=1}^N$ are the dataset scenarios. Note that with this choice of discrepancy, the $e$-th model-data constraint in \Cref{eq:generalcm} becomes
\begin{equation}
L_e \leq M_e(x) - c_e \leq U_e.
\end{equation}
This is exactly a version of the vector consistency measure presented in earlier work \cite[Equation (4.4)]{hegde2018consistency}. Other choices of $f$, such as the integral over $\mathcal{H}_c$ of the squared discrepancy (or its squared derivatives, should each $\Phi_i$ be differentiable), can also be handled in the B2BDC framework. 

Consistency measures formulated in this fashion gauge the disagreement between models and observations based on the ``simplest'' (or least complex) discrepancy required to render a dataset consistent. One potential application of this type of consistency measure is for model comparison. For a fixed set of basis functions $\{\Phi_i\}_{i=1}^{n}$, multiple models can be compared by evaluating \Cref{eq:generalcm}. This generalized consistency analysis with model discrepancy is currently being investigated and will be discussed in future work.
\section{Conclusions}
\label{sec:conclusion}
We examined the inclusion of model discrepancy as a linear combination of finite basis functions in B2BDC. The existing B2BDC framework was extended by reformulating the feasible set to include both model parameters and discrepancy-function coefficients; the prediction formulas were adjusted accordingly. Dataset consistency can be effectively recovered with the developed framework by increasing the complexity of the used model discrepancy function. The developed method offers a flexible construction of discrepancy function structure through the selection of basis functions; prior information on model discrepancy can be included naturally in the optimization problems as additional constraints. The confounding between model parameters and model discrepancy function in the posterior uncertainty, presented and discussed in statistical methodologies (e.g., \cite{brynjarsdottir2014learning}), was demonstrated from the deterministic perspective.

\newpage
\appendix
\section{Detailed and reduced hydrogen mechanisms} \label{append:mechanisms}
The detailed and reduced mechanisms used in the present work are listed in \Cref{tab:detailed H2 mechanism}. The reduced mechanism is consisted of the 5 reactions, marked with bold font and check marks.
\begin{table}[!htb]
\caption{Detailed and reduced H\textsubscript{2}-O\textsubscript{2} reaction sets and associated parameters of the rate coefficients, $AT^ne^{-E/RT}$, in the units of cm$^3$, mol, s, cal, K (from \cite{you2012process}). }\label{tab:detailed H2 mechanism}
\begin{minipage}{\textwidth}
\begin{center}
\begin{tabular}{c l l c c c} \hline
Reduced &  & \multicolumn{1}{c}{Reactions} & $A$ & $n$ & $E$ \\ \hline
$\text{\rlap{$\checkmark$}}$&1  & \textbf{H + O}$_\mathbf{2}$ \textbf{= O + OH} & $2.65\times 10^{16}$ & -0.6707 & 17041  \\ 
$\text{\rlap{$\checkmark$}}$&2  & \textbf{O + H}$_\mathbf{2}$ \textbf{= H + OH} & $3.87\times 10^{4}$ & 2.7 & 6260  \\
$\text{\rlap{$\checkmark$}}$&3  & \textbf{OH + H}$_\mathbf{2}$ = \textbf{H + H}$_\mathbf{2}$\textbf{O}  &  $2.16\times 10^{8}$ & 1.51 & 3430   \\
&4  & OH + OH = O + H$_2$O  &  $ 3.57\times 10^{4}$ & 2.4 & -2110   \\
&5\footnote{Collision efficiency: Ar = 0.63.}  & H + H + M = H$_2$ + M  &  $ 1.00\times 10^{18}$ & -1.0 &  0  \\
&  & H + H + H$_2$ = H$_2$ + H$_2$  &  $ 9.00\times 10^{16}$ & -0.6 &  0  \\
&  & H + H + H$_2$O = H$_2$ + H$_2$O  &  $ 6.00\times 10^{19}$ & -1.25 &  0  \\
&6\footnote{Collision efficiencies: H$_2$ = 2.4, H$_2$O = 15.4, Ar = 0.83.}  & O + O + M = O$_2$ + M  &  $1.20 \times 10^{17}$ & -1.0 &  0  \\
&7\footnote{\label{d}Collision efficiencies: H$_2$ = 2, H$_2$O = 12, Ar = 0.7.}  & O + H + M = OH + M  &  $ 4.71 \times 10^{18}$ & -1.0 &  0  \\
&8\footnote{Collision efficiencies: H$_2$ = 0.73, H$_2$O = 3.65, Ar = 0.38.}  & H + OH + M = H$_2$O + M &  $ 2.20\times 10^{22}$ & -2.0 &   0 \\
$\text{\rlap{$\checkmark$}}$&9\footnote{Collision efficiencies: H$_2$O = 12, Ar = 0.53; Troe parameters: $a$ = $0.5$, $T^{***}$ = $10^{-30}$, $T^*$ = $10^{30}$, $T^{**}$ = $10^{100}$.}  & \textbf{H + O}$_\mathbf{2}$ \textbf{+ M = HO}$_\mathbf{2}$\textbf{ + M}  &  $ 5.75\times 10^{19}$ & -1.4 &  0  \\
&  &  \textbf{H + O}$_\mathbf{2}$ \textbf{= HO}$_\mathbf{2}$ &  $ 4.65\times 10^{12}$ & 0.44 &  0  \\
&10  & H + HO$_2$ = O + H$_2$O  &  $ 3.97\times 10^{12}$ & 0.0 &  671  \\
$\text{\rlap{$\checkmark$}}$&11  & \textbf{H + HO}$_\mathbf{2}$ \textbf{= H}$_\mathbf{2}$ \textbf{+ O}$_\mathbf{2}$  &  $ 2.99\times 10^{6}$ & 2.12 &  -1172  \\
&12  & H + HO$_2$ = OH + OH  &  $ 8.40\times 10^{13}$ & 0.0 &  635  \\
&13  & O + HO$_2$ = OH + O$_2$  &  $ 2.00\times 10^{13}$ & 0.0 &   0 \\
&14  & OH + HO$_2$ = H$_2$O + O$_2$  &  $ 2.89\times 10^{13}$ & 0.0 &  -497  \\
&15  & HO$_2$ + HO$_2$ = H$_2$O$_2$ + O$_2$  &  $ 1.30\times 10^{11}$ & 0.0 & -1630   \\
&  & HO$_2$ + HO$_2$ = H$_2$O$_2$ + O$_2$  &  $ 4.20\times 10^{14}$ & 0.0 &  12000  \\
&16\footnote{Collision efficiencies: H$_2$ = 2, H$_2$O = 6, Ar = 0.67; Troe parameters: $a$ = $1.0$, $T^{***}$ = $10^{-30}$, $T^*$ = $10^{30}$, $T^{**}$ = $10^{30}$.}  & OH + OH + M = H$_2$O$_2$ + M  &  $ 1.46\times 10^{11}$ & 0.868 &  -8548  \\
&	& OH + OH = H$_2$O$_2$  &  $ 8.71\times 10^{9}$ & 0.869 & -2191   \\
&17	& H + H$_2$O$_2$ = H$_2$O + OH  &  $ 1.00\times 10^{13}$ & 0.0 &  3600  \\
&18  & H + H$_2$O$_2$ = HO$_2$ + H$_2$  &  $ 1.21\times 10^{7}$ & 2.0 &  5200  \\
&19  & O + H$_2$O$_2$ = HO$_2$ + OH  &  $ 9.63\times 10^{6}$ & 2.0 & 4000   \\
&20  & OH + H$_2$O$_2$ = H$_2$O + HO$_2$  &  $ 1.74\times 10^{12}$ & 0.0  & 318   \\
& & OH + H$_2$O$_2$ = H$_2$O + HO$_2$  &  $ 7.59\times 10^{13}$ & 0.0  & 7272  \\
&21\footref{d}  & O + OH + M = HO$_2$ + M  &  $ 8.00\times 10^{15}$ & 0.0 &   0 \\ \hline
\end{tabular}
\end{center}
\end{minipage}
\end{table}

\newpage

\bibliographystyle{siamplain}
\bibliography{references}
\end{document}